\documentclass[preprint,12pt]{elsarticle}

\usepackage{amssymb}

\usepackage{lineno}
\usepackage{multirow}
\usepackage{halloweenmath}
\usepackage{relsize}
\usepackage{listings}
\usepackage[]{algorithm2e}
\usepackage[hidelinks]{hyperref}

\journal{Nuclear Instruments and Methods in Physics Research - A}

\begin{document}
\begin{frontmatter}



\title{The CCube reconstruction algorithm for the SoLid experiment}


\author[inst2]{Y. Abreu}
\author[inst10]{Y. Amhis}
\author[inst3]{L. Arnold}
\author[inst8]{G. Barber}
\author[inst2]{W. Beaumont} 
\author[inst1]{S. Binet}
\author[inst9]{I. Bolognino}
\author[inst10]{M. Bongrand} 
\author[inst8]{J. Borg}
\author[inst10]{D. Boursette} 
\author[inst5]{V. Buridon}
\author[inst11]{B. C. Castle} 
\author[inst1]{H. Chanal} 
\author[inst3]{K. Clark}
\author[inst12]{B. Coupé}
\author[inst1]{P. Crochet}
\author[inst3]{D. Cussans}
\author[inst2,inst6]{A. De Roeck} 
\author[inst5]{D. Durand}
\author[inst13]{T. Durkin}
\author[inst9]{M. Fallot}
\author[inst8]{D. Galbinski}
\author[inst5]{S. Gallego}
\author[inst12]{L. Ghys}
\author[inst9]{L. Giot}
\author[inst8]{K. Graves}
\author[inst5]{B. Guillon} 
\author[inst9]{D. Henaff}
\author[inst8]{B. Hosseini} 
\author[inst10]{S. Jenzer}
\author[inst12]{S. Kalcheva}
\author[inst4]{L.N. Kalousis}
\author[inst4]{R. Keloth}
\author[inst11,inst15]{L. Koch}
\author[inst7]{M. Labare}
\author[inst5]{G. Lehaut} 
\author[inst3]{S. Manley}
\author[inst10]{L. Manzanillas}
\author[inst12]{J. Mermans}
\author[inst7]{I. Michiels} 
\author[inst1]{S. Monteil}
\author[inst7,inst12]{C. Moortgat}  
\author[inst3,inst13]{D. Newbold} 
\author[inst5]{V. Pestel}
\author[inst3]{K. Petridis}
\author[inst2]{I. Piñera}
\author[inst12]{L. Popescu} 
\author[inst10]{N. Roy}
\author[inst7]{D. Ryckbosch} 
\author[inst11]{N. Ryder}
\author[inst8]{D. Saunders}
\author[inst10]{M.-H. Schune} 
\author[inst2]{H. Rejeb Sfar}
\author[inst10,inst14]{L. Simard}
\author[inst8]{A. Vacheret}
\author[inst7]{G. Vandierendonck}
\author[inst12]{S. Van Dyck}
\author[inst4]{P. Van Mulders}
\author[inst2]{N. van Remortel}
\author[inst2,inst4]{S. Vercaemer}
\author[inst2]{M. Verstraeten}
\author[inst9]{B. Viaud}
\author[inst11,inst15]{A. Weber}
\author[inst1]{M. Yeresko}
\author[inst9]{F. Yermia}

\affiliation[inst1]{
organization={Université Clermont Auvergne, CNRS/IN2P3, LPCA},
city={Clermont-Ferrand},
country={France}
}

\affiliation[inst2]{
organization={Universiteit Antwerpen},
city={Antwerpen},
country={Belgium}
}

\affiliation[inst3]{
organization={University of Bristol},
city={Bristol},
country={United Kingdom}
}

\affiliation[inst4]{
organization={Vrĳe Universiteit Brussel},
city={Brussel},
country={Belgium}
}

\affiliation[inst5]{
organization={Normandie Univ, ENSICAEN, UNICAEN, CNRS/IN2P3, LPC Caen},
city={Caen},
country={France}
}

\affiliation[inst6]{
organization={CERN},
city={1211 Geneva 23},
country={Switzerland}
}

\affiliation[inst7]{
organization={Universiteit Gent},
city={Gent},
country={Belgium}
}

\affiliation[inst8]{
organization={Imperial College London, Department of Physics},
city={London},
country={}
}

\affiliation[inst9]{
organization={SUBATECH, Nantes Université, IMT Atlantique, CNRS/IN2P3},
city={Nantes},
country={France}
}

\affiliation[inst10]{
organization={IJCLab, Univ Paris-Sud, CNRS/IN2P3, Université Paris-Saclay},
city={Orsay},
country={France}
}

\affiliation[inst11]{
organization={University of Oxford},
city={Oxford},
country={United Kingdom}
}

\affiliation[inst12]{
organization={SCK-CEN, Belgian Nuclear Research Centre},
city={Mol},
country={Belgium}
}

\affiliation[inst13]{
organization={STFC, Rutherford Appleton Laboratory, Harwell Oxford, and Daresbury Laboratory},
city={Warrington},
country={United
Kingdom}
}

\affiliation[inst14]{
organization={Institut Universitaire de France},
city={Paris},
country={France}
}

\affiliation[inst15]{
organization={Johannes Gutenberg-Universität Mainz},
city={Mainz},
country={Germany}
}

\begin{abstract}

The SoLid experiment is a very-short-baseline experiment aimed at searching for nuclear-reactor-produced active-to-sterile antineutrino oscillations. The detection principle is based on the pairing of two types of \textit{solid} scintillators: polyvinyl toluene and $^6$LiF:ZnS(Ag), which is a new technology used in this field of Physics. In addition to good neutron-gamma discrimination, this setup allows the detector to be highly segmented (the basic detection unit is a 5~cm side cube). High segmentation provides numerous advantages, including the precise location of inverse beta decay (IBD) products, the derivation of the antineutrino energy estimator based on the isolated positron energy, and a powerful background reduction tool based on the topological signature of the signal. Finally, the system is read out by a network of wavelength-shifting (WLS) fibres coupled to a photodetectors. This paper describes the design of the reconstruction algorithm that allows maximum use of the granularity of the detector. The goal of the algorithm is to convert the output of the optical-fibre readout to the list of the detection units from which it originated. This paper provides a performance comparison for three methods and concludes with a choice of the baseline approach for the experiment.

\end{abstract}



\begin{keyword}
SoLid detector \sep positron signal reconstruction \sep ML-EM method
\PACS 0000 \sep 1111
\MSC 0000 \sep 1111

$\copyright$ 2023 by SoLid collaboration is licensed under \href{https://creativecommons.org/licenses/by/4.0/?ref=chooser-v1}{CC BY 4.0}
\end{keyword}

\end{frontmatter}


\newpage
\section{Introduction}
\label{sec:introduction}

The SoLid experiment is located in the vicinity of the BR2 research reactor at the SCK CEN site in Mol, Belgium. The experiment aims to make a precise measurement of the reactor antineutrino flux at a very short baseline (6.7 - 9.2 m), as well as its energy spectrum to study the ``5 MeV bump" ~\cite{DoubleChooz:2019qbj}. The first measurement allows, in principle, information on the so-called reactor anomaly RAA~\cite{Mention_2011} and the Gallium anomaly~\cite{Barinov_2022} to be provided. In particular, it is possible to test the 3+1 model~\cite{Abazajian:2012ys}, which assumes the existence of an additional light sterile neutrino state. The SoLid experiment was designed to probe the best-fit region of the oscillation parameters with sin($2\theta_s$) $\approx$ 0.1 and $\Delta$m$^2_s$ = 1~eV$^2$.

The description of the design of the SoLid detector is beyond the scope of this article. The interested reader can consult the detailed discussion reported in \cite{SoLid:2020cen}. The following paragraph contains the executive summary required for the description of the reconstruction algorithm. 
 
\begin{figure}[ht!]
    \centering
    \includegraphics[width=0.8\textwidth]{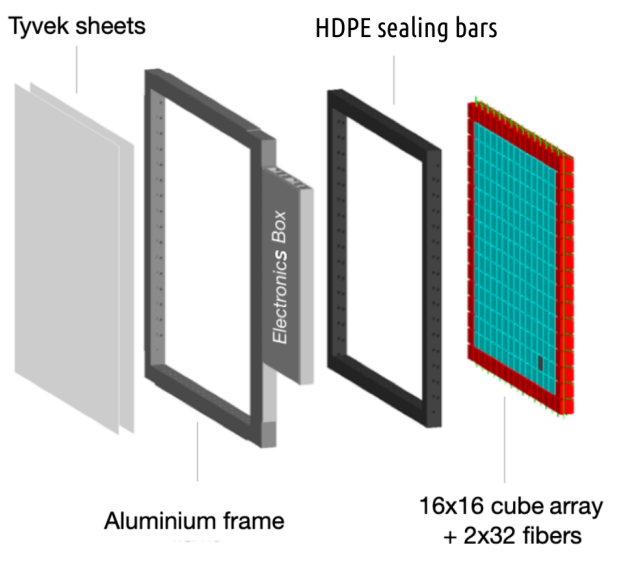}
    \caption{The composition of the SoLid detection plane.}
    \label{fig:solid_plane_module}
\end{figure}

The basic detection unit of the detector is a 5~cm side polyvinyl toluene (PVT) cube. PVT is a plastic scintillator that acts as a proton-enriched target for the antineutrino and detects the light generated by the IBD positron and consequent annihilation gamma. In addition, each cube has two neutron detection screens placed on adjacent faces. Each screen is a microcomposite $^6$LiF:ZnS(Ag). $^6$Li is responsible for neutron capture through the reaction in which the Li atom breaks down into $_2^4\alpha$ and $_1^3$H. Both products of the reaction are subsequently producing the scintillation light in the ZnS microcrystals. The detection units are combined into planes of 16$\times$16 units each. Each cube is individually wrapped in Tyvek to prevent scintillation light from escaping. Furthermore, each plane is surrounded by two square Tyvek sheets, which optically decouple the planes by preventing light from passing between them even further (shown in Figure~\ref{fig:solid_plane_module}). The latter is a very important feature of the calibration algorithm. Ten planes make up a module and there are five modules in total. The scintillation light from the PVT cube is collected by two vertical and two horizontal WLS fibres that pass through each detection cell. One side of each WLS fibre is covered with a Mylar foil that acts as a mirror to reflect the incoming light. The second side is coupled with the multi-pixel photon counters (MPPCs) that read out the light. The digitised version of the readout received from the detector is the starting point for any further analysis. Therefore, signal candidates must be defined from this input. The most convenient and useful way to represent the data is to transform the MPPC response back into the list of detection cubes involved in the event. This type of problem, \textit{i.e.}\@ image reconstruction, has been investigated in medical physics, for example, in \cite{med_image_reco}. 

Several algorithms have been applied for the reconstruction of SoLid data. Two of them are already well-known and widely used. The first is a regularisation approach called the Fast Iterative Shrinkage Threshold Algorithm, or FISTA, which is an improved version of ISTA~\cite{fista}. The second is the Bayesian Maximum-Likelihood Expectation-Minimisation (ML-EM) approach, which is used \textit{e.g.}\@ in the NEXT experiment~\cite{next}. Finally, the last method is custom-made and builds on the ML-EM algorithm by augmenting it with information from the underlying physics processes at work. All three of them will be presented in this article and their performances will be compared.

Note that all the results presented in this paper are obtained solely with Monte Carlo (MC) simulations. The production of simulation files in the SoLid experiment is based on several components. First, a detailed simulation of the core of the BR2 reactor is produced. It couples the reference ILL $\beta$-spectrum, fission rate predictions extracted from the MCNPX 3D model of the reactor core combined with the CINDER90 evolution code, and the MURE code to track the burn-up of the fissile products. Second, a detailed \textsc{Geant4}~\cite{Geant4:2002zbu} simulation of the geometry of the SoLid detector is performed. Third, the SoLid Oscillation (SoLO) framework, which works as an antineutrino generator and mediator between the two simulations introduced above. For each signal event, SoLO picks a position inside the reactor core according to the given fission map, an antineutrino energy from the predicted spectrum, and the interaction point inside the detector. By construction, this approach takes into account the geometrical acceptance of the detector. Thus, the computational cost is significantly reduced compared to isotropic generation. To reduce it even further, SoLO directly transfers the kinematic information of the positron and neutron to the detector simulation. The interested reader can consult the detailed description of the SoLO framework reported in~\cite{Ianthe}. The final piece that handles the output of the \textsc{Geant4} simulation is the so-called ReadOut Simulation (ROSim). The goal of ROSim is to reproduce the response from the detector. Practically, it transforms the energy deposits into the number of scintillation photons that are read out by the correspondent MPPCs. The detailed tuning of the ROSim parameters is reported in~\cite{maja}.
\section{Signal definition}
\label{sec:signal_definition}

The SoLid detector uses the IBD process ($\bar{\nu}_e + p \to n + e^+$) to detect antineutrinos. This justifies the use of two scintillators: $^6$LiF:ZnS(Ag) screens to capture the neutron and PVT as a proton-enriched target for the antineutrino, hence recording the positron scintillation. For the former, once the neutron is thermalised through elastic collisions, it is captured and produces tritium and $\alpha$ particles, detected in the ZnS(Ag) crystal which is integrated in the neutron-detection screens. The latter detects in addition to the positron scintillation the subsequent positron annihilation gammas. 

Two main sources of background can mimic the signal: the natural radioactivity of the detector, \textit{e.g.} bismuth-polonium (BiPo) cascade, and the atmospheric neutrons creating spallations in and around the detector. The former is strongly suppressed by using the properties of scintillation in ZnS detector crystals. The latter is reduced in the SoLid detector by maximally using the granularity of the PVT cells, and hence relies on the cube reconstruction method discussed in this article.

The antineutrino energy is defined from the IBD process as follows:

\begin{equation}
\begin{gathered}
    E_{\bar{\nu}} + m_p = E_{e^+} + m_{e^+} + E_n + m_n\\
    E_{\bar{\nu}} = E_{e^+} + m_{e^+} + m_n - m_p = E_{e^+} + 1.806 \: \textnormal{MeV} \; ,
\end{gathered}
\label{eq:e_estimator}
\end{equation}

\noindent where $m_p$ and $m_n$ denote the masses of the scattered proton and the outgoing neutron, respectively, and $E_i$ with $i \in {e^+,n, \bar{\nu}}$. Neglecting $E_n$ in the second equation is justified, since the neutron kinetic energies do not exceed 50 keV, much lower than the $\mathcal{O}$(3)~MeV antineutrino energy. 

Therefore, the antineutrino energy estimator relies on the measurement of the actual energy deposited by the positron, in contrast to the total prompt energy of the event in the case of liquid scintillators. The size of the cube in the SoLid geometry corresponds to the maximum path length of a 10~MeV positron. According to Geant4 studies, the positron indeed deposits its energy in a single cube in 80\% of events. Furthermore, the cube in which the annihilation occurred (Annihilation Cube or AC) is the most energetic cube of the event when the positron energy is above 1~MeV. Thus, an accurate reconstruction of the AC is key for a precise measurement of antineutrino energy. 

The first attempt to determine the AC was based on the calculation of the sum of the digitised output of the MPPCs for each plane. The cubes were then placed at each position where the two vertical and two horizontal fibres intersect, and the AC cube, as any other cube, was given the sum of the four fibres as its energy. Signal selection consisted of taking the highest-energy cube alone in this plane. Several penalties are affecting this strategy: removal of the events with several cubes in the same plane; creation of fake cubes (which will be called ghosts in the following); non-separation of the annihilation gammas. Incidentally, signal events were overwhelmed by background sources.

Therefore, a more educated reconstruction algorithm is desirable. A solution that would provide a list of all cubes where the physics interaction took place in the event has several advantages: no need to restrict the analysis to events with a single cube in the plane;   the events in which the positron deposits energy in two cubes can also be used (15\% of the statistics); cubes containing the annihilation gamma energy deposits can be used to select the signal. The final point allows powerful discriminative features of the signal signature to be designed to suppress background contamination. If both annihilation gamma clusters are reconstructed, the requirement that they are found back-to-back in the detector is a highly efficient background rejection tool. Consequently, the energy estimator in Equation~\ref{eq:e_estimator} is correctly defined. The advantages listed above do not only motivate the search for an adequate reconstruction algorithm, but form the core of the \textit{topological}\@ analysis of the SoLid Phase I data, which will be used in future physics publications. Of course, keeping the level of ghosts (fake cubes) under control will be a cornerstone of the reconstruction method to design. 
\section{The CCube algorithm}
\label{sec:ccube}

\begin{figure}[ht!]
    \centering
    \includegraphics[width=0.5\textwidth]{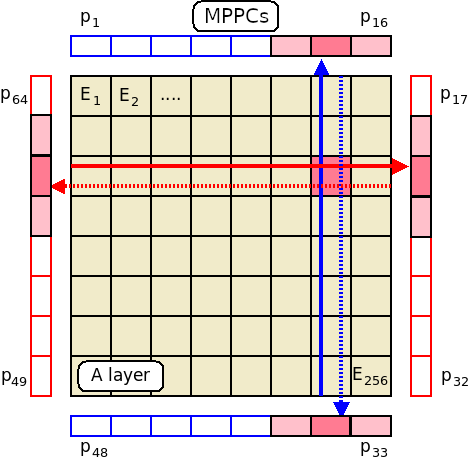}
    \caption{A sketch of the energy deposit (pink cube) in the SoLid detector plane with fired horizontal (red) and vertical (blue) fibres and impacted (pink and light pink) MPPC.}
    \label{fig:ccube}
    \end{figure}

The purpose of the CCube algorithm (which stands for Clermont-Cube) is to perform a reverse engineering procedure and to transform the list of digitised MPPC readout into a list of cubes where physics interactions took place. Communication between the detection cubes and the MPPC is provided by the WLS fibres. Therefore, the fibres act as projectors from the cubes to the electronics. The projection is assumed to be linear with respect to the deposited energy. The detector planes are optically decoupled in the SoLid design, which means that only cubes in the same plane as the triggered fibres are considered. This simplifies the reconstruction problem from 3D to 2D. Finally, the CCube algorithm assesses the energy of a cube if and only if signals are received in at least one horizontal and one vertical direction. The typical number of cubes in a physics event is less than twenty. Therefore, the reconstruction algorithm deals only with a small subset of detector planes and within a plane with a small number of fibres. Hence, the selection of the algorithm to test is not \textit{a priori} limited by the CPU consumption constraint. However, optimisation of computational cost is further addressed in the design of the custom approach. An example of the reconstruction problem is sketched in Figure~\ref{fig:ccube}.

Let us mathematically define the problem. By postulating a$_{ij}$ as the projector from cube \textit{j}\@ to MPPC \textit{i}\@, the readout value is calculated as follows:

\begin{equation}
    a_{i,1}\cdot E_{1} + a_{i,2}\cdot E_2 \;+\; ... \;+\; a_{i,256}\cdot E_{256} \;=\; p_i \; ,
\end{equation}

\noindent where 256 corresponds to the number of cubes in a plane. The number of projections is twice the sum of the number of cubes in rows and columns. It makes 64 projections and hence 64 equations for the SoLid geometry. These equations can be represented in a compact matrix form as: 

\begin{equation}
\label{eq:sm}
    AE = p \;,
\end{equation}

\noindent where $p$ is the column vector of the readout projections. It is important to note that the element $p_i$ corresponds to the number of photo avalanches that the MPPC receives. $E$ is the column vector of unknown energy deposits to be determined by the reconstruction procedure and $A$ is a matrix of 64 $\times$ 256 dimensions. Matrix A is called the System Matrix and embodies the best of our knowledge about the detector behaviour at each stage, from the light generation to the digitisation, and the absolute energy scale. It is derived from the sample of cosmic horizontal muons with a statistical precision below 1\%. The method of derivation is beyond the scope of this article and will be described in a future paper. For now, it is assumed that all elements of the matrix, together with the absolute energy scale, can be determined. 

\begin{figure}[ht!]
    \centering
    \includegraphics[width=0.75\textwidth]{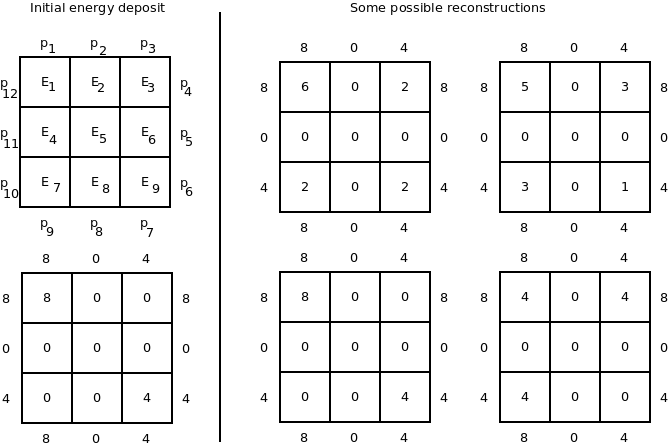}
    \caption{An illustration that the reconstruction problem is ill-posed. The initial digitised readout interpreted in multiple ways in the sense of the attached energy contributions.}
    \label{fig:ghost1}
    \end{figure}

Equation~\ref{eq:sm} is a discrete linear inverse problem. The problem is well-posed if it meets certain requirements: existence, uniqueness, and stability of the solution with respect to the small shifts. In the current case, the uniqueness condition is violated, as illustrated in Figure~\ref{fig:ghost1}. For simplicity, each cube communicates there only through the fibres that cross it: the projectors a$_{ij}$ of the cubes crossed by the given fibre i are set to 1, and to 0 for all other pairs of i,j.
This indicates that only two energy deposits in the same plane are enough to allow the readout to be fulfilled in several ways. In particular, false cubes may be created at the intersection of the fibres of the true ones. In addition to the multiplicity of events, false cubes also introduce an energy bias to the initial energy deposits. Thus, the reconstruction problem is ill-posed and shall be solved either by regularisation or Bayesian methods. Before discussing the methods themselves, let us define the set of estimators that are used to assess the performance of different methods.

\section{Reconstruction procedure estimators}
\label{sec:estimators_ccube}

The estimators used to evaluate the performance of the various methods are divided into two categories. The first is related only to the convergence of the algorithms. The estimators in this group calculate the deviation (also indicated as \textit{data fidelity})\@ of the solution derived (E) from the actual measurements (p). The least squares norm L2 and the Kullback-Leiber divergence KL~\cite{10.1214/aoms/1177729694} were selected to obtain it. Their mathematical definitions read as follows:

\begin{gather}
\label{eq:norms}
    T_{L2}(E) \; = \; \|AE \; - \; p\|_2^2,\\
    T_{K\!L}(E) \; = \; AE - p + p \cdot \log p - p \cdot \log{AE} \; .
\end{gather}

\noindent With these two estimators, an immediate cross-check can be performed and the behaviour outliers can be tracked. Moreover, the usage of both metrics allows the different physical models of the measurement process to be tested ($L2$ prefers a Gaussian model, while $K\!L$ prefers a Poisson model). 

The second group of estimators involves physical performance. \textsc{Geant4} (G4) simulation serves as truth-level information, while the output of the SoLid Analysis Framework (Saffron2 ~\cite{peste2019} or simply S2) with the reconstruction approach at work serves as a derived solution. The comparison relies on the list of cubes per event, their coordinates (positions in the plane) and associated energy contributions. To begin with, the total number of cubes in which the energy deposit occurs is tracked at each level: 

\begin{itemize}
    \item n$_{\textnormal{cubes}}^{G4}$ represents the number of cubes at the \textsc{Geant4} level,
    \item n$_{\textnormal{cubes}}^{S2}$ represents the number of cubes at the Saffron2 level.
\end{itemize}

The reconstruction efficiency is defined from the ratio of cubes present in both lists. If a cube is present in the Saffron2 list and not in the \textsc{Geant4} list, it is counted as a ghost in the quantity $n^{S2}_{\textnormal{cubes}}(\overline{G4})$. Reconstruction efficiency and ghost rate are therefore computed, respectively, as: 

    \begin{equation}
    \label{eq:cube_eff}
        \mathlarger{\epsilon}_{\rm{RECO}} \;=\; \frac{n^{G4}_{\textnormal{cubes}}(S2)}{n_{\textnormal{cubes}}^{G4}} \cdot 100\%.
    \end{equation}
    
    \begin{equation}
    \label{eq:ghost_rate}
    G_{\textnormal{RATE}} \;=\; \mathghost \;=\; \frac{n^{S2}_{\textnormal{cubes}}(\overline{G4})}{n_{\textnormal{cubes}}^{S2}} \cdot 100\% \;.    \end{equation}

\noindent The number of unreconstructed cubes is an energy-dependent parameter. The lower the energy of a given cube, the higher the probability that it will not be reconstructed. The situation is similar to that of ghost cubes, which generally have very low energies. To assess this dependence, the physics estimators are evaluated for several different energy thresholds. The threshold removes all cube energy deposits at the levels of Saffron2 and \textsc{Geant4}, which have energy below the threshold. 
\section{The regularisation approach}
\label{sec:regularisation}

This method does not aim at explicitly solving the initial equation, but rather to obtain a well-posed problem, similar to the starting ill-posed one, and solve it instead. Therefore, the solution obtained is an approximation. An example of such an approach is Tikhonov regularisation ($T\!R$)~\cite{tikhonov1977solutions}:

\begin{equation}
\label{eq:reg}
    E_{T\!R} \;=\; \underset{E}{\textnormal{min}} \; \{\|AE \; - \; p\|_2^2 + \lambda \| LE \| ^2_2 \} \; ,
\end{equation}

\noindent where the second part of the expression is known as a regularisation term. In this term, $L$ is a Tikhonov matrix and $\lambda$ is the regularisation parameter, respectively. In other words, the expression $A^{-1}p$ (assuming that $A$ is an invertible square matrix) does not satisfy Equation~\ref{eq:sm}. The level of deviation is controlled by the regularisation parameter $\lambda$. This is the penalty to obtain a unique solution. In Equation~\ref{eq:reg} the regularisation term is quadratic; its choice strongly depends on the nature of the initial problem. It is known~\cite{4407762, 1217267} that for sparse problems the so-called $l_1$ approach can be employed: it uses the absolute values of the target value (energies) as the regularisation term. The low multiplicity of cubes per plane makes the SoLid reconstruction a canonical sparse problem. Therefore, the loss function reads: 

\begin{equation}
\label{eq:reg2}
    E_{L1} \;=\; \underset{E}{\textnormal{min}} \; \{\|AE \; - \; p\|_2^2 + \lambda \| E \|_1 \} \; .
\end{equation}

\noindent The most popular methods for solving such equations are gradient-based, since they significantly decrease the required computational resources. Gradient-based algorithms, in general, use the iterative approach, which relies on improving the current knowledge of the solution with additional information provided by the gradient of the initial function. For example, in the iterative shrinkage-thresholding algorithms (ISTA), the iterative step is completed by the shrinkage-soft-threshold step, which provides the following: 

\begin{equation}
\begin{gathered}
    E_n \;=\; T_{\lambda \eta} (E_{n-1} \;-\; \eta\cdot\nabla F(E_{n-1})) \\
    T_\alpha(x) \;=\; sgn(x)\cdot(|x| - \alpha)_+ \; ,
\end{gathered}
\end{equation}

\noindent where $\eta$ is the step size, \textit{sgn}\@ is a sign function and $T$ is a shrinkage operator. Recently, an upgraded version of the ISTA algorithm was developed. It has an improved complexity result of $\mathcal{O}$(1/k$^2$), which is achieved by using the specific linear combination of the two previous iteration values within the shrinkage operator. It is called FISTA~\cite{fista} (for Fast ISTA) and was implemented in Saffron2. 

\begin{figure}[ht!]
    \centering
    \includegraphics[width=0.475\textwidth]{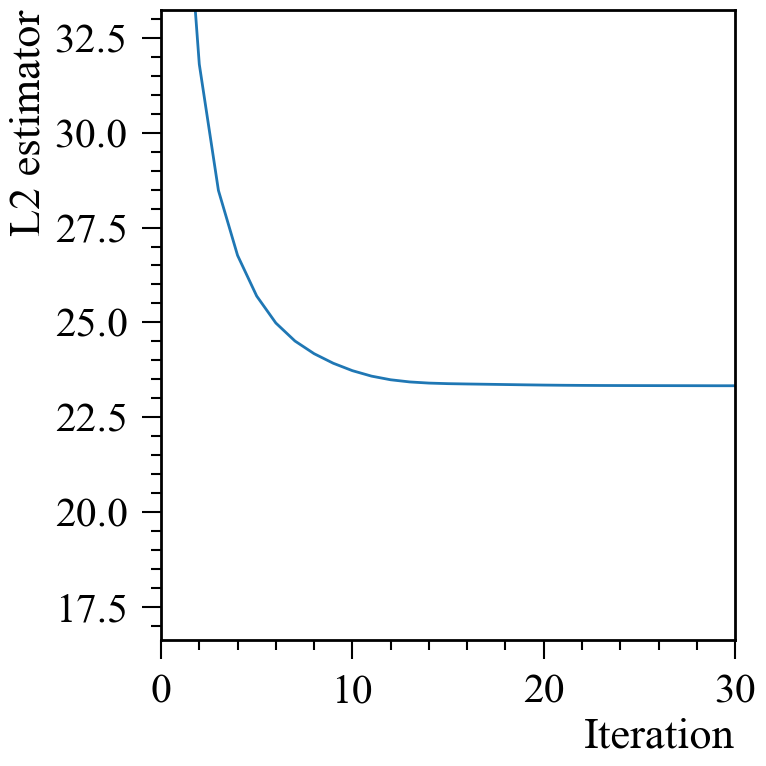}
    \includegraphics[width=0.475\textwidth]{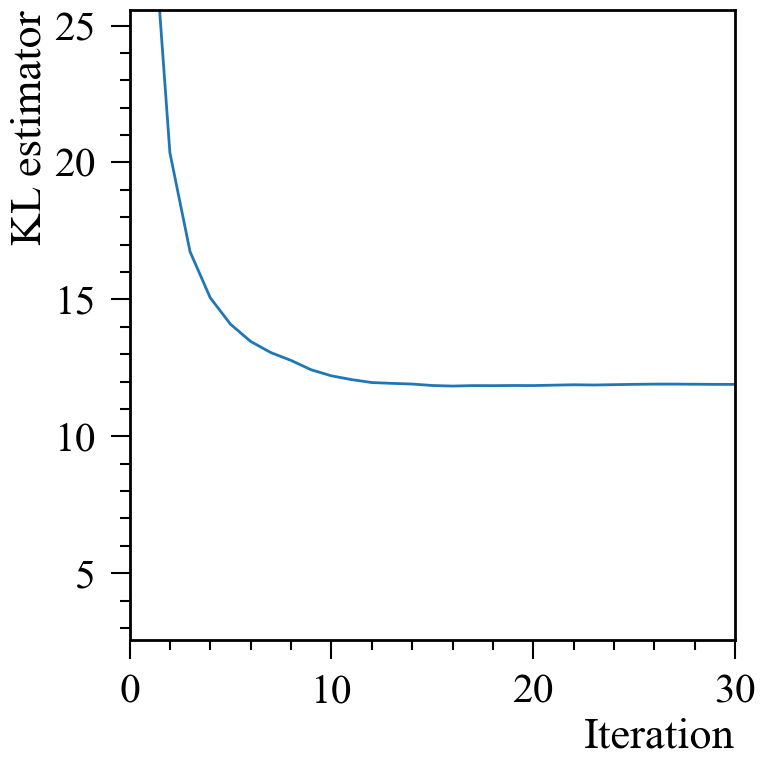}
    \caption{Data fidelity value for the FISTA algorithm versus the number of executed iterations. The distribution is obtained from 3$\times$10$^5$ simulated IBD events for L2 and KL estimators. The results for both of them are in agreement.}
    \label{fig:conv_fista}
\end{figure}

The method was tested with a sample of 3$\times$10$^5$ simulated IBD events. The convergence development is presented in Figure~\ref{fig:conv_fista}. With either of the calculation methods, convergence improves with the number of iterations and reaches a plateau. The high value of data fidelity is explained by the fact that it is the sum of all events generated. The number of iterations was arbitrarily selected as thirty, and the physics estimators were obtained after this condition. They are summarised in Table~\ref{tab:physics_fista} for several energy thresholds. As expected, the ghost-cube rate decreases drastically with larger energy cut values. However, a compromise has to be made in the choice of the energy threshold, since the reconstruction efficiency of the IBD annihilation gammas is decreasing accordingly. Dedicated optimisation must be performed in the analysis workflow, which is beyond the scope of this paper. Together with the reduction of the multiplicity of ghosts, the reconstruction efficiency improves with a larger threshold. This is another illustration that the algorithm copes much more easily with high-energy cubes.

\begin{table}[ht!]
    \centering
    \begin{tabular}{c|cccccccccc}
    Cut (keV) & 25 & 50 & 75 & 100 & 125 &  150 & 175 & 200 & 225 & 250 \\
    \hline\hline
    $\mathghost$ (\%)& 27.2 & 20.7 & 15.8 & 12.4 & 9.6 & 7.6 & 6.2 & 5.0 & 4.1 & 3.4\\
    $\mathlarger{\epsilon}$ (\%) & 61.2 & 66.1 & 71.0 & 75.1 & 78.3 & 81.0 & 83.4 & 85.4 & 87.2 & 88.6\\
    \end{tabular}
    \caption{The ghost cube rate and the cube reconstruction efficiency for the FISTA algorithm, obtained from simulated IBD events processed with Saffron2.}
    \label{tab:physics_fista}
\end{table}

\section{The Bayesian approach}
\label{sec:bayesian}

Bayesian statistics offers an alternative approach to regularisation. It will not provide a unique solution and the output is dependent on the initial guess. However, it presents advantageous attributes that are discussed in this section. The method \textit{per se}\@ is based on the assumption that both energy deposits \textit{E}\@ and the measured projections \textit{p}\@ are stochastic variables with the following characteristics:

\begin{itemize}
    \item $\pi_{\textnormal{data}}(p|E)$ is a probability density function (p.d.f.), which describes the likelihood of measuring $p$ given the knowledge of $E$;
    \item $\pi_{\textnormal{prior}}(E)$ describes the nature of the considered signal;
    \item $\pi_{\textnormal{posterior}}(E|p)$ is a posterior distribution, \textit{i.e.}\@ the p.d.f., which describes the probability to have $E$ given $p$.
\end{itemize}

By considering the Bayes master formula, the following relation between the three is obtained: 

\begin{equation}
\pi_{\textnormal{posterior}}(E|p) \;\propto\; \pi_{\textnormal{prior}}(E) \cdot \pi_{\textnormal{data}}(p|E) \; ,
\end{equation}

\noindent where $\propto$ denotes proportionality. In this approach, $\pi_{\textnormal{posterior}}(E|p)$ is the solution of Equation~\ref{eq:sm}. The ML-EM method has the following set of features: accurate modelling of the Poisson noise distribution; the solution is constrained to be positive without regularisation; and unitarity of the sum of the projections is preserved (the total number of scintillation photons is conserved). The measured projections in SoLid obey the Poisson law. Therefore, the selection of the Maximum-Likelihood Expectation-Minimisation (ML-EM) method is one of the most adapted. It is widely used in medical imaging and recently in particle physics, \textit{e.g.}\@ in the NEXT experiment~\cite{next}.

The algorithm iteratively generates the projected values based on the current approximation of the solution. Iterations are stopped at the desired precision. An initial guess of the answer is required at the beginning of the process. The choice of initial guess is critical because of the potential non-uniqueness of the solution; it is desirable to find the globally optimal solution. Two methods have attempted to choose a starting point. First, a {\it democratic} approach that assumes an equal share of the total energy between the cubes. Second, the FISTA result described above is used. It has been shown~\cite{Zhu:18} that the choice of the latter significantly improves, in general, the performance of ML-EM for sparse problems, and this is also observed in this work. Equation~\ref{eq:emml} defines the master formula of the ML-EM algorithm. It is obtained by maximising the probability of obtaining a measurement \textit{p}\@.

\begin{equation}
\label{eq:emml}
    E_{j}^{n+1} \;=\; \frac{E_{j}^{n}}{\sum\limits_i a_{ij}}\sum_i a_{ij} \frac{p_i}{\sum\limits_{k}a_{ik}E_{k}^{n}} \;
\end{equation}  

\begin{table}[ht!]
    \centering
    \begin{tabular}{c|cccccccccc}
    Cut (keV)& 25 & 50 & 75 & 100 & 125 & 150 & 175 & 200 & 225 & 250\\
    \hline\hline
    $\mathghost$ (\%)& 25.9 & 19.3 & 13.0 & 9.5 & 7.0 & 5.3 & 4.2 & 3.4 & 2.7 & 2.3\\
    $\mathlarger{\epsilon}$ (\%) & 61.5 & 66.9 & 72.6 & 77.2 & 80.8 & 83.5 & 85.9 &  87.9 & 89.4 & 90.6 \\
    \end{tabular}
    \caption{The ghost cubes rate and the cube reconstruction efficiency for the FISTA+ML-EM algorithm, obtained from the simulated IBD events processed with Saffron2.}
    \label{tab:physics_fista_mlem}
\end{table}

\begin{figure}[ht!]
    \centering
    \includegraphics[width=0.475\textwidth]{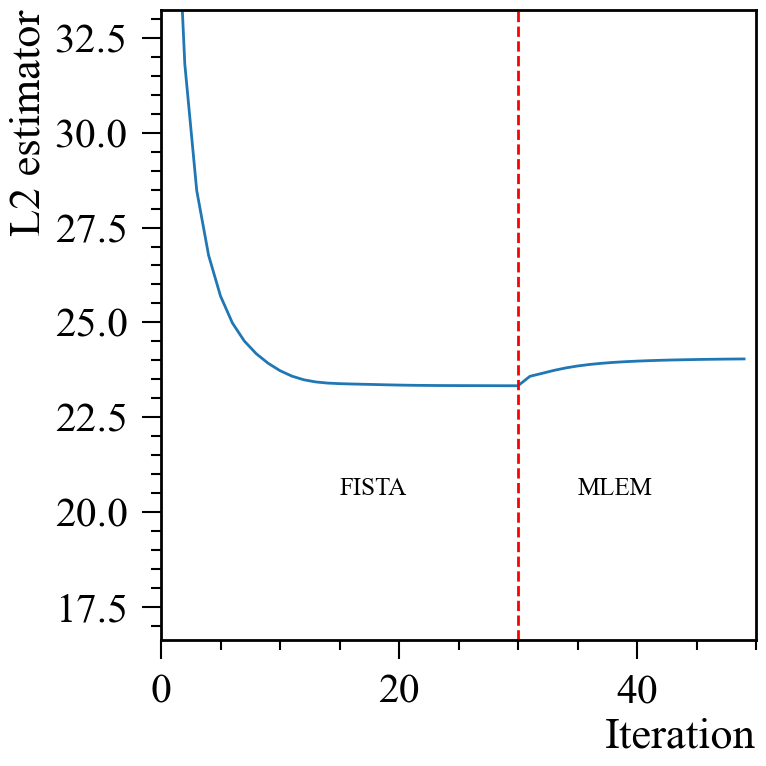}
    \includegraphics[width=0.475\textwidth]{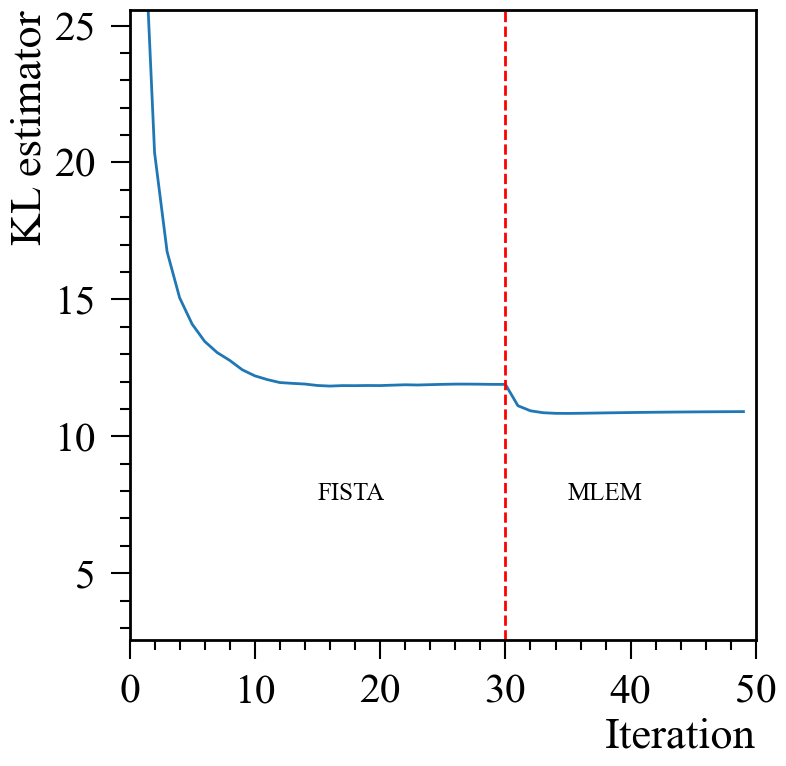}
    \caption{Data fidelity value for the FISTA + ML-EM algorithm versus the number of executed iterations. It is obtained from 3$\times$10$^5$ simulated IBD events for L2 and KL estimators.}
    \label{fig:conv_fista_mlem}
\end{figure}

\begin{figure}[ht!]
    \centering
    \includegraphics[width=0.35\textwidth]{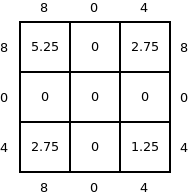}
    \hspace{2 cm}
    \includegraphics[width=0.35\textwidth]{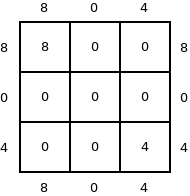}
    \caption{The two possible solutions to the simplified reconstruction problem posed in Figure~\ref{fig:ghost1}, which are equally favoured by the ML-EM approach.}
    \label{fig:fista_init_problem}
\end{figure}

The combined FISTA and ML-EM method was implemented and tested using the same MC IBD sample as for the regularisation approach. Thirty iterations of FISTA are followed by 15 iterations of ML-EM. The physics estimators are summarised in Table~\ref{tab:physics_fista_mlem}. The overall trend with respect to the cut value highlighted in the previous subsections remains the same for both observables. However, this composite approach improves overall performance for all the thresholds considered. The most interesting change appears in the data fidelity distributions displayed in Figure~\ref{fig:conv_fista_mlem}. The KL data fidelity is expectedly improved after ML-EM takes over. Taking into account the physics features of the performed measurement (\textit{i.e.}\@ its Poisson nature) improves the performance of the reconstruction algorithm. At the same time, the algorithm cannot claim which of the two configurations presented in Figure~\ref{fig:fista_init_problem} is better. 

\section{The SoLid approach}
\label{sec:solid}

IBD events exhibit two characteristics that are useful for reconstruction. They are sparse, and the energy deposits from positron ionisation are larger than the two from the annihilation gamma ones. The former has to be reconstructed more accurately, since it is in the core of the antineutrino energy estimator. Therefore, starting with the assignment of energy to the most energetic cube before the others, an improvement in the performance of the reconstruction algorithm is expected. This improvement includes the decrease in the ghost cube rate and lift of the degeneracy introduced in Figure~\ref{fig:fista_init_problem}. 

The Orthogonal Matching Pursuit (OMP) algorithm fits this objective. At each stage, the aim is to find the best energy deposit that optimises the L2 norm (see Equation~\ref{eq:norms}). The choice of the L2 norm over the KL norm is motivated by the simplicity of the implementation and the fact that it prioritises the most energetic cube in the event. Due to the higher energies and thus a larger number of photo avalanches, the Poisson behaviour is less prominent for these cubes. The method first adjusts the positron energy deposit before moving on to smaller energy deposits. Like FISTA, OMP will be used as the initialiser of the ML-EM algorithm. In this case, the full capacity of the OMP is not required. To reduce computational cost, a simplified OMP (sOMP) is derived. The main difference between the two is that the simplified version allows one to assign the energy to a cube only once. Therefore, once a cube is accessed, it is removed from the list of options. Hence, the sOMP algorithm is run once and is used as an initialiser of ML-EM. 

\begin{figure}[ht!]
    \centering
    \includegraphics[width=0.475\textwidth]{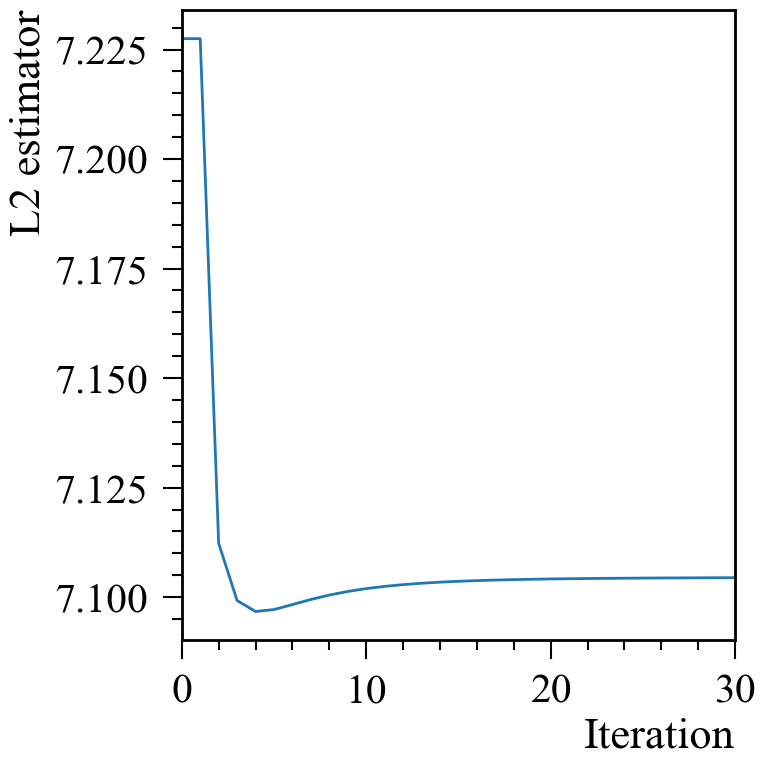}
    \includegraphics[width=0.475\textwidth]{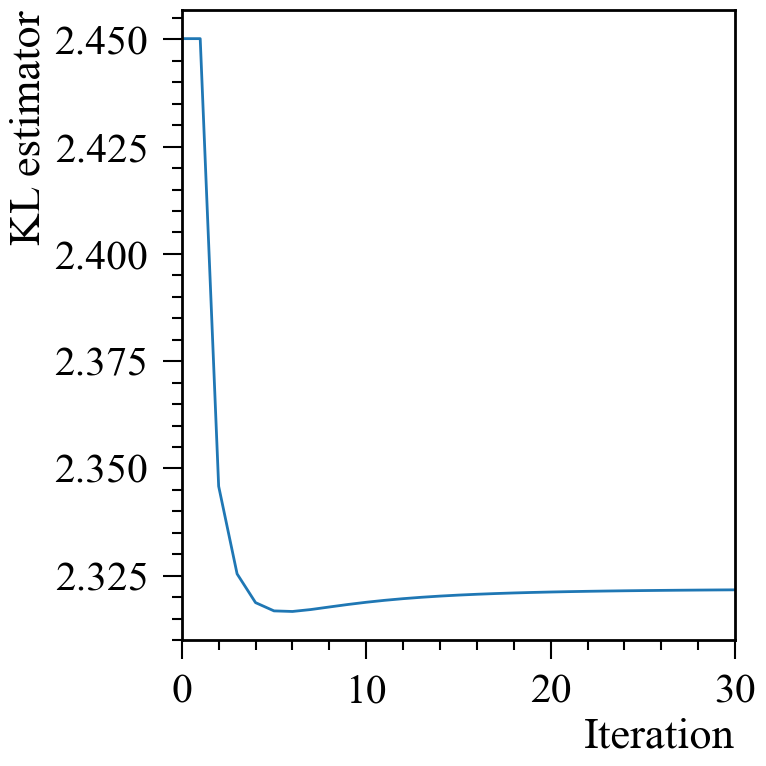}
    \caption{Data-fidelity value for the sOMP+ML-EM algorithm versus the number of executed iterations. The function is obtained from 3$\times$10$^5$ simulated IBD events for L2 and KL estimators.}
    \label{fig:conv_somp_mlem}
\end{figure}

The combined sOMP and ML-EM approach is assessed on the same IBD sample as the other methods. The data-fidelity distributions displayed in Figure~\ref{fig:conv_somp_mlem} show that this method converges faster than those described in the previous sections. The value of the data-fidelity estimator at the inflection point is similar to the value obtained at the plateau, also yielding similar performance for the reconstruction. Therefore, thirty iterations have been adopted in the actual implementation of the SoLid reconstruction. An improvement is observed for the physics estimators summarised in Table~\ref{tab:physics_somp_mlem}. For an energy threshold of 75 keV, the ghost rate is reduced by one-third for a comparable cube-reconstruction efficiency. A similar performance enhancement is observed for all energy cut values. 

\begin{table}[ht!]
    \centering
    \begin{tabular}{c|cccccccccc}
    Cut (keV)& 25 & 50 & 75 & 100 & 125 & 150 & 175 & 200 & 225 & 250\\
    \hline\hline
    $\mathghost$ (\%) & 22.4 & 15.0 & 9.6 & 6.5 & 4.5 & 3.4 & 2.5 & 1.9 & 1.5 & 1.2 \\
    $\mathlarger{\epsilon}$ (\%) & 55.6 & 63.8 & 71.2 & 76.8 & 80.9 & 84.0 & 86.7 & 88.6 & 90.3 & 91.5 \\
    \end{tabular}
    \caption{The ghost cubes rate and the cube reconstruction efficiency for the sOMP+ML-EM algorithm, obtained from the simulated IBD events processed with Saffron2.}
    \label{tab:physics_somp_mlem}
\end{table}

The final comparison between the three methods is summarised in Table~\ref{tab:physics_relative}. The results presented are given for an energy threshold of 75~keV, which is the actual choice of the reconstruction adopted in SoLid.

\begin{table}[ht!]
    \centering
    \begin{tabular}{c|ccc}
    Method & FISTA & FISTA+ML-EM & sOMP+ML-EM\\
    \hline \hline
    $\mathghost$ (\%) & 15.8 & 13.0 & 9.6\\
    $\mathlarger{\epsilon}$ (\%) & 71.0 & 72.6 & 71.2\\
    \end{tabular}
    \caption{The performance comparison of the three reconstruction methods.}
    \label{tab:physics_relative}
\end{table}

\subsection{Energy resolution}
\label{sec:energy_resolution}

The combined sOMP and ML-EM approach grants the best reconstruction performance. It is interesting to check that the energy resolution complies with the expectations of the detector design. This study is carried out using events selected by the so-called topological algorithm: empty 3$\times$3$\times$3 envelope around the most energetic cube (AC)  in the event and two additional electromagnetic clusters reconstructed outside that envelope. Event candidates are also required to have the most energetic cube above 1.5~MeV. These conditions ensure that the AC candidate meets the requirements of Equation \ref{eq:e_estimator}. Therefore, the energy spread is defined as: 

\begin{equation}
        E_{\textnormal{SPREAD}} \; = \; \frac{(E_{\textnormal{TRUE}} - 1.806) - E_{\textnormal{RECO}}}{(E_{\textnormal{TRUE}} - 1.806)} \; = \; \frac{(E_{G4} - 1.806) - E_{S2}}{(E_{G4} - 1.806)} \;, 
\end{equation}

\noindent where $E_{\textnormal{G4}}$ is the antineutrino energy from the \textsc{Geant4} simulation and $E_{S2}$ is the energy of the cube with the highest energy in the ROSim event. Events in the sample defined above are split into 25 bins with identical statistics. For each bin, the $E_{\textnormal{SPREAD}}$ distribution is built and fit with a Crystal Ball function~\cite{crystalball}. The fit result provides a value of $\sigma$, which is used to compute $\sigma_E/E$. In this case, $E$ is the median energy for each individual bin. The energy resolution studies for the same type of scintillator (EJ-200) are presented in~\cite{ej200resolution}. It has been chosen to adopt the same opportunistic 3-parameter function to fit the data:   

\begin{equation}
    \frac{\sigma_{E}}{E} = \sqrt{C_2 + \frac{C_1}{E} + \frac{C_0}{E^2}}.
\end{equation}

\begin{figure}[ht!]
    \centering
    \includegraphics[width=1.\textwidth]{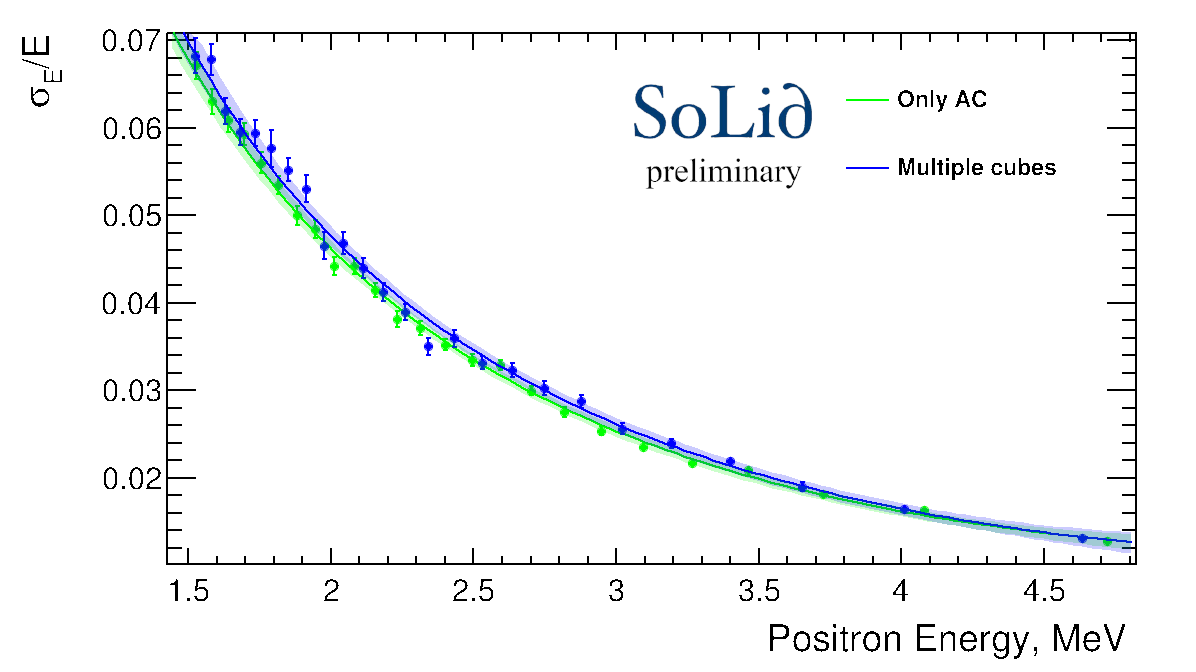}
    \caption{Energy resolution as obtained from the energy estimator of ~\ref{eq:e_estimator}. The green curve corresponds to AC candidate alone in the plane and the blue curve corresponds to several cubes in the same plane.}
    \label{fig:energy_resolution}
\end{figure}

\noindent Figure~\ref{fig:energy_resolution} shows the fit of this function to the energy resolution simulation points. It is observed that the same energy resolution is obtained for events with only one cube in the plane (trivial case as a reference) and with several cubes in the same plane (the application case of the algorithm). Detector design performance is not altered by the reconstruction algorithm as far as the energy measurement is concerned.     

\subsection{Open dataset performance}
\label{subsec:open_perf}

The reconstruction algorithm presented in this article and its performance have been developed and obtained using simulated data. The performance of the reconstruction algorithm on the data can be illustrated by a search for antineutrino signal in the SoLid detector. The reactor-off data sample and an open dataset are used (\textit{i.e.}\@ processed with the SoLid analysis software and the CCube reconstruction algorithm in particular) for this purpose. Reactor-on samples are recorded during periods when the BR2 reactor operates at nominal power. Hence, such samples contain both signal and background events. The open data set (denoted ROn) is a small reactor-on sample of 21 days recorded during a period of 19 June to 10 July 2018. Reactor-off samples are recorded during periods when the BR2 reactor is off, respectively. These samples are used to study background sources and model their properties. The reactor off data sample (ROff) used for this study is the sum of all reactor-off periods recorded during SoLid Phase I (mid-2018 to mid-2020). The IBD signal model is determined with simulation data. The sum of the signal and background models is then compared to the open data set.

The aim of the CCube algorithm is to reconstruct the position of a signal candidate cube and the energy deposit associated with it. The cube where positron annihilation occurs contributes the most to the antineutrino energy estimator. Therefore, the variables that evaluate the performance of the reconstruction are the energy of the annihilation cube itself (AC energy) and the spatial distance in cubes between the annihilation cube and the neutron capture cube ($\Delta$R, which serves as a measurement of the precision of position assignment).

The level of background in the raw SoLid data is overwhelming (the signal-to-background ratio is at the level of 1 to 1000). Antineutrino candidates are searched for in events where multiple cubes are present. This ensures that there is enough information to separate annihilation photons and positron signals. The CCube algorithm is instrumental in the separation of the annihilation gammas and positron ionisation components of the electromagnetic signal. This \textit{topological}\@ selection is followed by multivariate selections that take advantage of the physical characteristics of the signal candidates that can be built from cube information. For example, the angle between the momentum vectors of the annihilation gamma, built from the annihilation gamma and the positron cubes; a back-to-back $\pi$ radian angle will more likely designate a signal candidate.

\begin{figure}[ht!]
    \centering
    \includegraphics[width=1\textwidth]{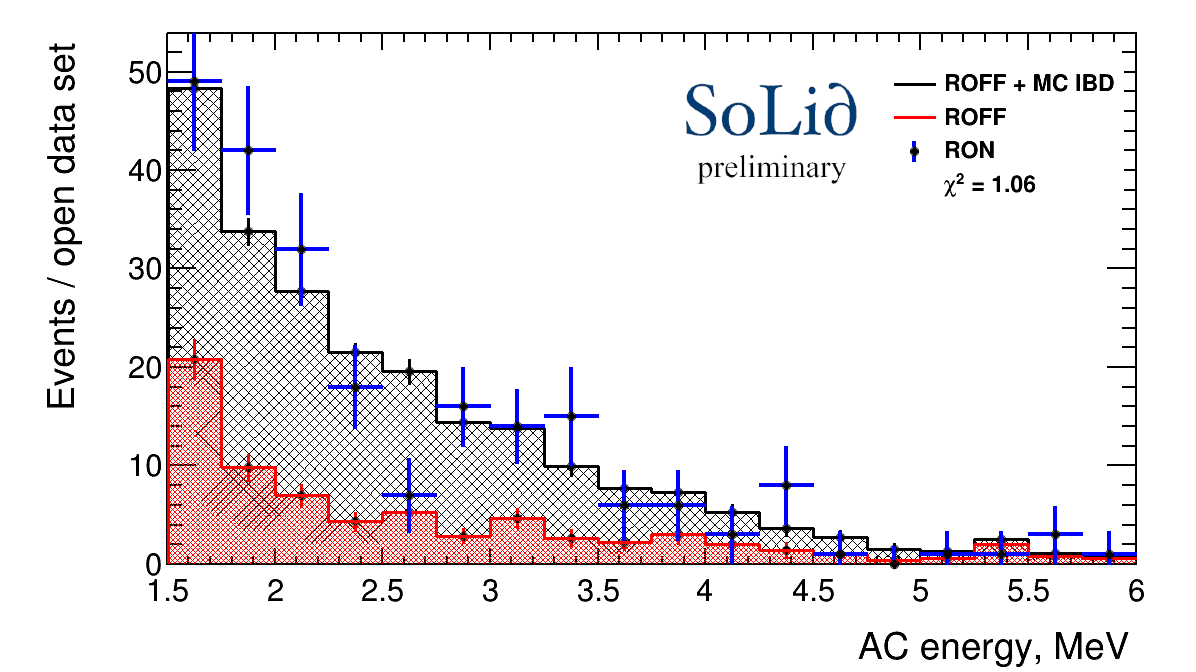}
    \includegraphics[width=1\textwidth]{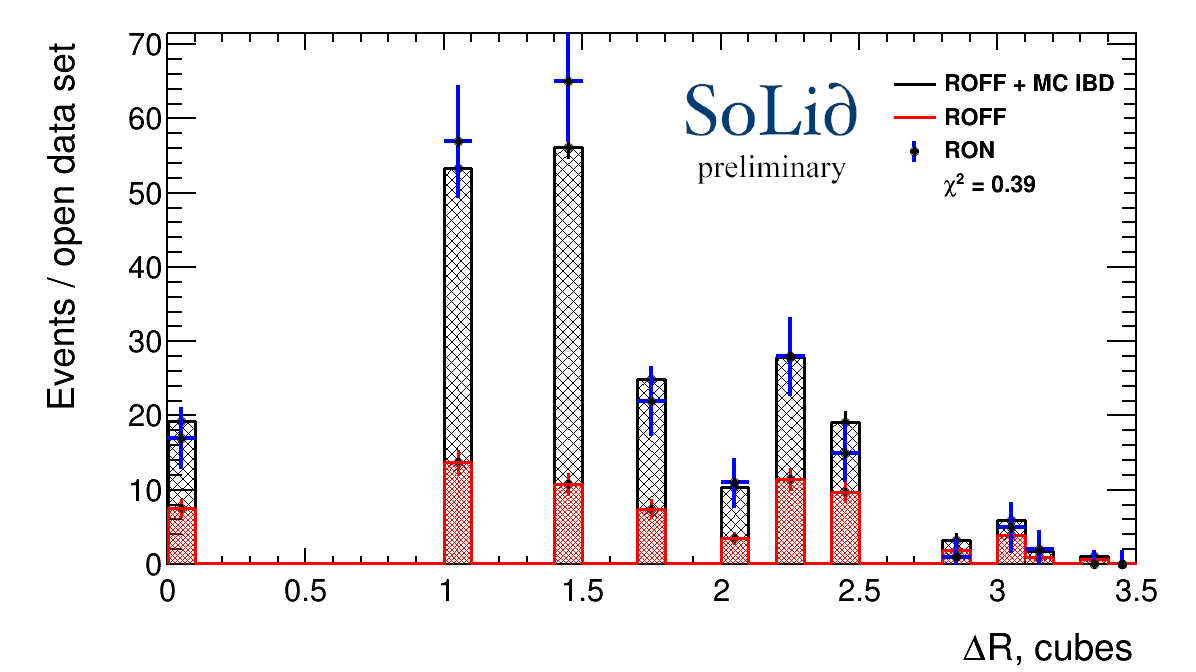}
    \caption{The comparison between the predicted background (red) and signal (grey) model to the open data set (blue data points) for the annihilation cube energy (top) and the spatial distance between annihilation cube and neutron capture cube (bottom). Their agreement, measured by a $\chi^2$ test, is satisfactory.}
    \label{fig:control_5d_topo20}
\end{figure}

Figure~\ref{fig:control_5d_topo20} shows the distributions of the energy of the positron candidates and their distance from the neutron-capture cube candidate in the ROn sample. The sum of the histograms of the background (red) and signal (grey) models agrees with the open data set (blue data points) within statistical uncertainty. There is a clear excess of events with respect to the ROff prediction, consistent with antineutrino IBD candidates. The reconstruction algorithm described in this article is primarily used to define the signal. It also proved to be essential in the definition of discriminative variables in order to successfully select the signal.
\section{Conclusion}
\label{sec:conclusion}

The CCube reconstruction algorithm is presented in this article. It is of central importance for characterising the electromagnetic signal clusters in the SoLid reconstruction and therefore serves as the cornerstone of the definition and selection of IBD candidates recorded by the SoLid experiment. Three different methods employing established algorithms are designed to solve this reconstruction problem. The sOMP+ML-EM algorithm, which benefits from the use of the underlying physics of IBD events, exhibits the best performance. The choice of the 75~keV threshold was dictated by the original energy threshold in the antineutrino selection analysis. The combined sOMP+ML-EM algorithm is considered the state-of-the-art method by the SoLid collaboration.
\section{Acknowledgements}
\label{sec:Acknowledgements}

This work was supported by the following funding agencies: Agence Nationale de la Recherche grant ANR-16CE31001803, Institut Carnot Mines, CNRS/IN2P3, and Region Pays de Loire, France; FWO-Vlaanderen and the Vlaamse Herculesstichting, Belgium; The UK groups acknowledge the support of the Science \& Technology Facilities Council (STFC), United Kingdom; We are grateful for the early support given by the subdepartment of Particle Physics at Oxford and High Energy Physics at Imperial College London. We also thank our colleagues, the administrative and technical staff of the SCK CEN, for their invaluable support for this project. Individuals have received support from FWO-Vlaanderen and the Belgian Federal Science Policy Office (BelSpo) under the IUAP network programme; the STFC Rutherford Fellowship programme and the European Research Council under the European Union Horizon 2020 Programme (H2020-CoG)/ERC Grant Agreement n. 682474; Merton College Oxford.

\newpage
\appendix

 \bibliographystyle{elsarticle-num} 
 \bibliography{cas-refs}





\end{document}